\documentclass
[aps,twocolumn,showpacs,superscriptaddress,tightenlines]{revtex4}
\usepackage{epsfig,rotating}
\usepackage{amsmath}
\usepackage{amsfonts}
\usepackage{amssymb}
\usepackage{graphicx}
\providecommand{\U}[1]{\protect\rule{.1in}{.1in}}
\providecommand{\U}[1]{\protect\rule{.1in}{.1in}}
\providecommand{\U}[1]{\protect\rule{.1in}{.1in}}
\begin{document}
\title{Pauli blocking and medium effects in nucleon knockout reactions}
\author{C.A. Bertulani}\email{carlos_bertulani@tamu-commerce.edu}
\affiliation{Physics Department, Texas A\&M University-Commerce,
Commerce, TX 75428-3011, USA}
\author{C. De Conti}\email{conti@itapeva.unesp.br}
\affiliation{Campus Experimental de Itapeva, Universidade Estadual Paulista, 18409-010, Itapeva, SP, Brazil}
\date{\today }

\begin{abstract}
We study  medium modifications of the
nucleon-nucleon (NN) cross sections and their influence on the nucleon knockout reactions. Using the eikonal approximation, we compare the results obtained with  
free NN cross sections 
with those obtained with a purely geometrical treatment of Pauli-blocking  and with NN obtained with more elaborated Dirac-Bruecker methods.  
The medium effects are parametrized in terms of the baryon density. We focus
on symmetric nuclear matter, although the geometrical Pauli-blocking also allows for the treatment of asymmetric nuclear 
matter. It is shown that medium effects can change the nucleon knockout cross sections and momentum distributions up to 10\% 
in the energy range $E_{lab}=50-300$ MeV/nucleon. The effect is more evident in reactions involving halo nuclei.\end{abstract}

\maketitle

\section{Introduction}

Nuclear structure calculations are now able
to  reproduce the measured masses, charge
radii and low-lying excited states of a large number of nuclei. For
very exotic nuclei, the small additional stability that comes with
the filling of a particular orbital can have profound effects upon
their existence as bound systems, their lifetime and structure.
The determination of the ordering, spacing and the occupancy of
orbitals is therefore essential in assessing how exotic nuclei evolve in the
presence of large neutron or proton excess and to what extent theories have
predictive power. Such spectroscopy of the single-particle structure in
short-lived nuclei typically uses direct nuclear reactions.

Nucleon knockout reactions  at intermediate
energies   have become a well-established and
quantitative tool for studying the location and occupancy of single-particle
states and correlation effects in the nuclear many-body system, as discussed in
Refs. \cite{BM92,Gregers,Tostevin1999,han03,Gade2008a}. In a peripheral, sudden
collision of the 
fast-moving projectile, a
single nucleon is removed from the projectile, producing projectile-like
residues  in the exit channel~\cite{han03}.
Referred to the center-of-mass
system of the projectile, the transferred momentum is ${\bf k}_c$. In the
sudden approximation and for the knockout reaction,  this must equal the momentum of the struck nucleon
before the collision. The measured partial cross-sections to individual
final levels provide spectroscopic factors for the individual
angular-momentum components $j$. In complete analogy
to the use of angular distributions in transfer reactions, the orbital
angular momentum $l$ is in the knockout reactions revealed
by the distributions of the quantity ${\bf k}_c$.

Extensions of the nucleon knockout formalism includind the treatment of
final-state interactions were discussed in Ref. \cite{BH04} where one has shown
that  Coulomb final-state interactions are of
relevance. They can be done by just adding the Coulomb phase
$\phi=\phi_N+\phi_C$ in the eikonal phase as described \cite{BH04}. Inclusion of higer-order effects \cite{AS00,AOS03} and a theory for
two-nucleon knockout  \cite{Tostevin2006,Sim09,Simpson2009} has been
developed. Knockout reactions represent a particular case for which higher
projectiles energies allow a simpler theoretical treatment of the reaction
mechanism, due to the simplicity of the eikonal scattering waves and the
assumption of a single-step process.      

A question of interest is the anti-symmetrization of the full projectile-target scattering wavefunction. At intermediate energies
($\sim 100$ MeV/nucleon), this effect is usually neglected. In the Glauber formalism of knockout reactions the scattering waves are calculated from an optical potential based on nucleon-nucleon scattering cross sections. A rough treatment of anti-symmetrization is obtained by the manifestation of medium modification of the nucleon-nucleon cross section. Knowledge of the
medium modification of the nucleon-nucleon (NN) cross section is necessary for an
adequate numerical modelling of heavy-ion collision dynamics in central collisions (see,
e.g. Ref.~\cite{Li08} and refs. therein). In these collisions, the ultimate purpose is to
extract information about the nuclear equation of state (EOS) by
studying global collective variables describing the collision
process. In direct reactions, such as knockout reactions, the medium effects on the NN cross sections are much smaller because mostly low nuclear densities are probed. The goal in this work is to identify if medium modifications of NN scattering modify appreciably the  cross sections in knockout reactions. A systematic study of this effect in the literature is still lacking and is the focal point of this article. 

Medium modifications of NN cross sections are usually treated within the Brueckner-Hartree-Fock (BHF) theory, where the $G$-matrix
serves as the in-medium scattering amplitude, with medium effects
being introduced through the self-consistent nuclear mean field
and Pauli blocking. The literature in this subject is very long, see e.g., Refs. \cite{LM:1993,LM:1994,FK06,Sam08}.
In addition to being a fundamental input for nuclear reactions at high energies, the in-medium cross sections provide an immediate
connection with the nucleon mean free path, $\lambda$, one of the
most fundamental quantities characterizing the propagation of
nucleons through matter. In turn, $\lambda$ enters the calculation
of the nuclear transparency function where the later is related to
the total reaction cross section, $\sigma_R$, of a nucleus \cite{HRB91}.

In this work we include  medium effects of the
NN cross section in knockout reactions with  a simple
geometrical treatment of Pauli-blocking and also with more elaborated Dirac-Brueckner results  
 in terms of baryon densities. We focus
specifically on symmetric nuclear matter.  We calculate knockout cross sections and momentum distributions for selected reactions. After the introductory remarks in this section, in
Section II we describe the formalism used in our
calculations. Section III contains our numerical results. We
conclude in Section IV with our summary.

\section{Knockout reactions}

\subsection{Medium modification of nucleon-nucleon cross sections}

\begin{figure}[!t]
\centering
\includegraphics[totalheight=8.0cm]{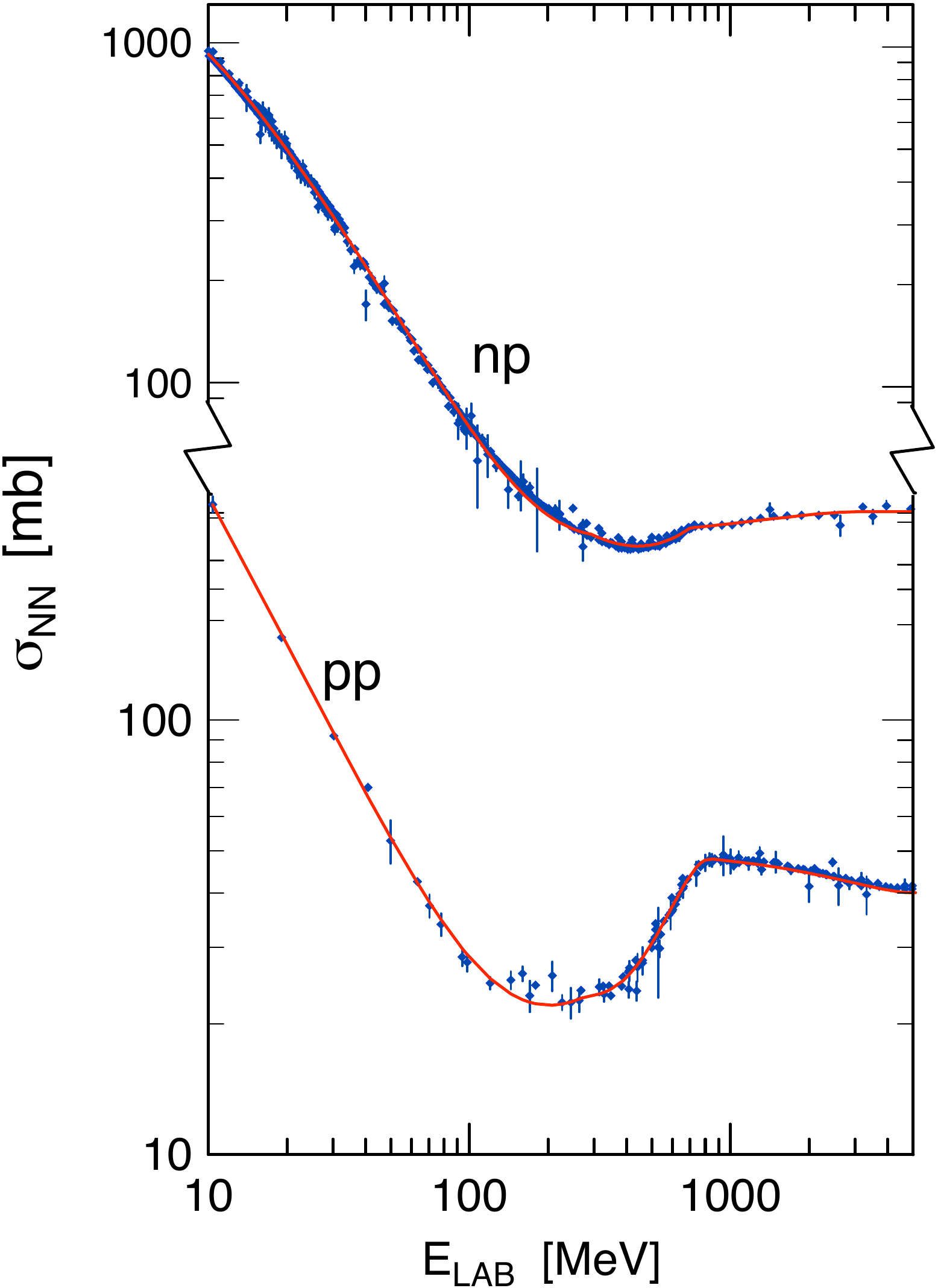}
\vspace{5mm} \caption{(Color online)  Least square fit (solid
curves)  to the nucleon-nucleon cross section described by  Eqs.
(\ref{signn1},\ref{signn2}). The experimental data are  from Ref.
\cite{pdgxnn}. } \label{signn}
\end{figure}

The free (total) nucleon-nucleon cross sections were taken from the
Particle Data Group  \cite{pdgxnn}. For our practical purposes, we have developed new fits for the free
nucleon-nucleon cross sections, separated in three energy intervals, by means of the expressions
\begin{equation}
\sigma_{pp}=
\left\{
\begin{array}
[c]{c}%
19.6+{4253/ E} -{ 375/ \sqrt{E}}+3.86\times 10^{-2}E \\
({\rm for }\ E < 280\  {\rm MeV}) \\ \; \\
32.7-5.52\times 10^{-2}E+3.53\times 10^{-7}E^3  \\
-  2.97\times 10^{-10}E^4  \\
({\rm for }\   280\ {\rm MeV}\le E < 840\  {\rm MeV}) \\ \; \\
50.9-3.8\times 10^{-3}E+2.78\times 10^{-7}E^2 \\
 +1.92\times 10^{-15} E^4  \\
({\rm for}\  840 \ {\rm MeV} \le E \le 5 \ {\rm GeV})\end{array}
\right.
\label{signn1}
\end{equation}
for proton-proton collisions, and
\begin{equation}
\sigma_{np}=
\left\{
\begin{array}
[c]{c}%
89.4-{2025/ \sqrt{E}}+{19108/ E}-{43535/ E^2}
\\
 ({\rm for }\ E < 300\  {\rm MeV}) \\ \; \\
14.2+{5436/ E}+3.72\times 10^{-5}E^2-7.55\times 10^{-9}E^3
 \\
 ({\rm for }\   300\ {\rm MeV}\le E < 700\  {\rm MeV}) \\ \; \\
33.9+6.1\times 10^{-3}E-1.55\times 10^{-6}E^2 \\
 +1.3\times 10^{-10}E^3\\
 ({\rm for}\  700 \ {\rm MeV} \le E \le 5 \ {\rm GeV}) \end{array}
\right.
\label{signn2}
\end{equation}
for proton-neutron collisions. $E$ is the projectile laboratory
energy. The coefficients in the above equations have been obtained
by a least square fit to the nucleon-nucleon cross section
experimental data over a variety of energies, ranging from 10 MeV to
5 GeV. In figure \ref{signn} these fits are represented by a solid
line whereas the filled circles are the experimental data from ref.
\cite{pdgxnn}.

Most practical studies of medium corrections of nucleon-nucleon scattering are done by considering  the effective two-nucleon 
interaction in infinite 
nuclear matter, or G-matrix,
as a solution of the Bethe-Goldstone equation \cite{GWW58}
\begin{eqnarray}
&&\langle\mathbf{k}|\mathrm{G}(\mathbf{P},\rho_1,\rho_2)|\mathbf{k}_{0}\rangle
=\langle\mathbf{k}|\mathrm{v}_{NN}|\mathbf{k}_{0}\rangle\nonumber \\
&-&\int{\frac
{d^{3}k^{\prime}}{(2\pi)^{3}}}{\frac{\langle\mathbf{k}|\mathrm{v}%
_{NN}|\mathbf{k^{\prime}}\rangle Q(\mathbf{k^{\prime}},\mathbf{P}%
,\rho_1,\rho_2)\langle\mathbf{k^{\prime}}|\mathrm{G}(\mathbf{P},\rho_1,\rho_2)|\mathbf{k}%
_{0}\rangle}{E(\mathbf{P},\mathbf{k^{\prime}})-E_{0}-i\epsilon}}\nonumber \\
\,\label{10}%
\end{eqnarray}
with $\mathbf{k}_{0}$, $\mathbf{k}$, and $\mathbf{k^{\prime}}$ the
initial, final, and intermediate relative momenta of the NN pair,
${\bf k}=({\bf k}_1-{\bf k}_2)/2$ and ${\bf P}=({\bf k}_1+{\bf k}_2)/2$. If energy and momentum is conserved in the binary collision,
${\bf P}$ is conserved in magnitude and direction, and the magnitude of ${\bf k}$ is also conserved. $\mathrm{v}_{NN}$ is the
nucleon-nucleon potential. $E$ is the energy of the two-nucleon
system, and $E_{0}$ is the same quantity on-shell. Thus
$
E(\mathbf{P},\mathbf{k})=e(\mathbf{P}+\mathbf{k})+e(\mathbf{P}-\mathbf{k}%
)$, with $e$ the single-particle energy in nuclear matter. It is also implicit in  Eq. \eqref{10} that the final momenta ${\bf k}$ of the NN-pair
 also lie outside the  range of occupied states. 

Eq.~(\ref{10}) is density-dependent due to the presence of the Pauli
projection operator $Q$, defined by
\begin{equation}
Q(\mathbf{k},\mathbf{P},\rho_1,\rho_2)=\left\{
\begin{array}
[c]{c}%
1,\ \ \ \mathrm{if}\ \ \ k_{1,2}>k_{F1,F2}\\
0,\ \ \ \ \ \mathrm{otherwise.}%
\end{array}
\right. \label{12}%
\end{equation}
with $k_{1,2}$ the magnitude of the momenta of each nucleon. $Q$
prevents scattering into occupied intermediate states. The Fermi momenta $k_{F1,F2}$ are related to the proton and neutron 
densities by means of the zero temperature density approximation, $k_{Fi}=(3\pi^2 \rho_i)^{1/3}$. For finite nuclei, one usually replaces $\rho_i$ by the local 
densities to obtain the local Fermi momenta. This is obviously a rough approximation, but very practical and extensively used in the literature.

\begin{figure}[!t]
\centering
\includegraphics[totalheight=8.0cm]{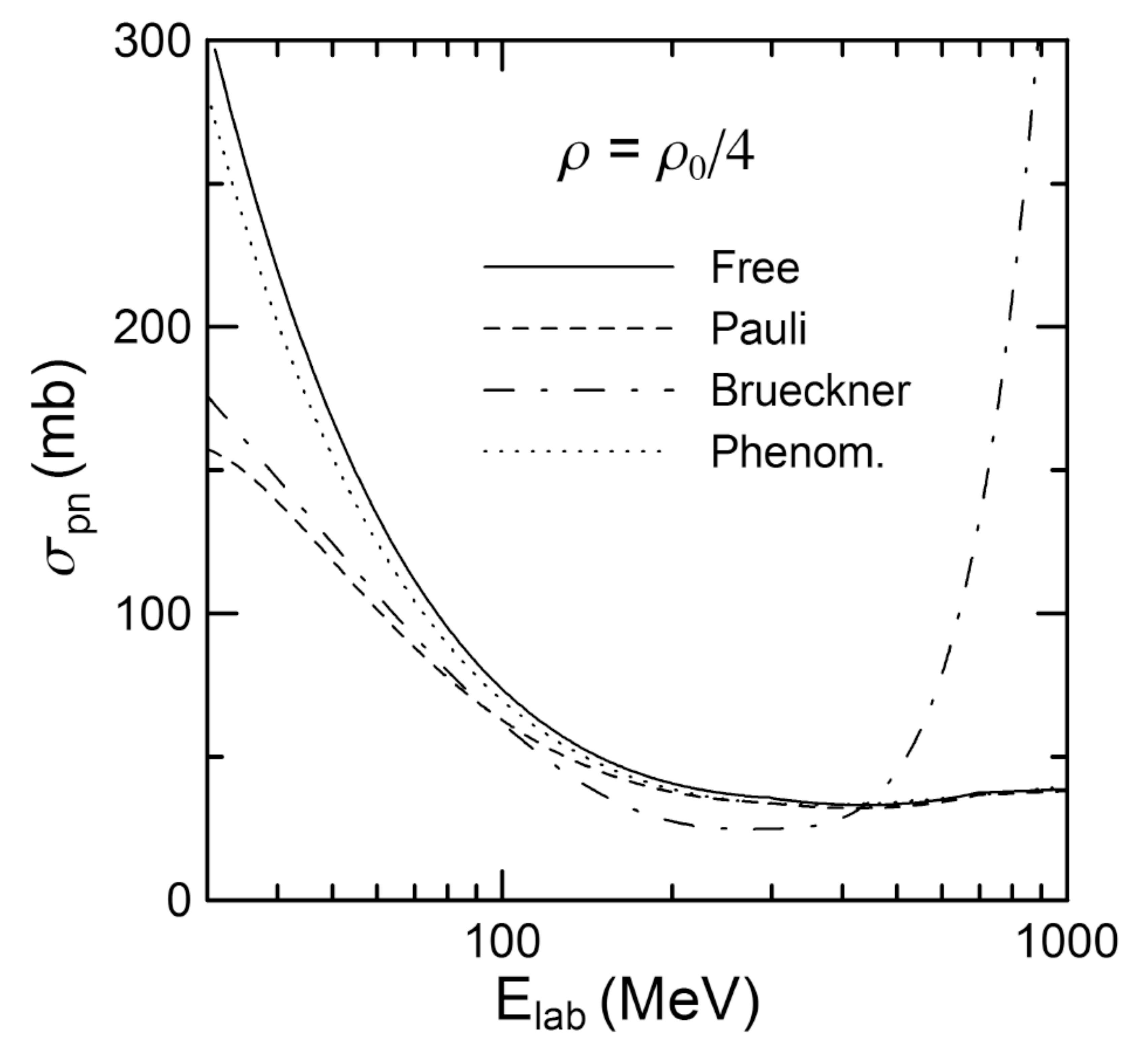}
\vspace{5mm} \caption{Parametrizations of proton-neutron cross sections as a function of the laboratory energy. The solid line is the parametrizarion  of the free $\sigma_{pn}$ cross section given by Eq. \eqref{signn2}.  The other curves include medium effects for symmetric nuclear matter for $\rho=\rho_0/4$, where $\rho_0=0.17$ fm$^{-3}$. The dashed curve includes the geometrical effects of Pauli blocking, as described by Eq.  \eqref{VM1}. The dashed-dotted curve is the result of the Brueckner theory, Eq. \eqref{brueckner}, and the dotted curve is the phenomenological parametrization, Eq. \eqref{pheno}.} \label{signpe}
\end{figure}

Only by means of several approximations, Eq. \eqref{10} can be related to nucleon-nucleon cross sections. If one neglects the medium 
modifications of the nucleon-mass, and scattering through intermediate states, the medium modification of the NN cross sections can be
accounted for by the geometrical factor $Q$ only, i.e.
\begin{equation}
\sigma_{NN}(k,\rho_1,\rho_2)=\int {d\sigma_{NN}^{free}\over d\Omega}  Q(k,P, \rho_1,\rho_2) d\Omega , \label{snngeo}
\end{equation}
where $Q$ is now a simplified geometrical condition on the available scattering angles for the scattering of the NN-pair to unoccupied final
states.

\begin{figure}[!t]
\centering
\includegraphics[totalheight=8.0cm]{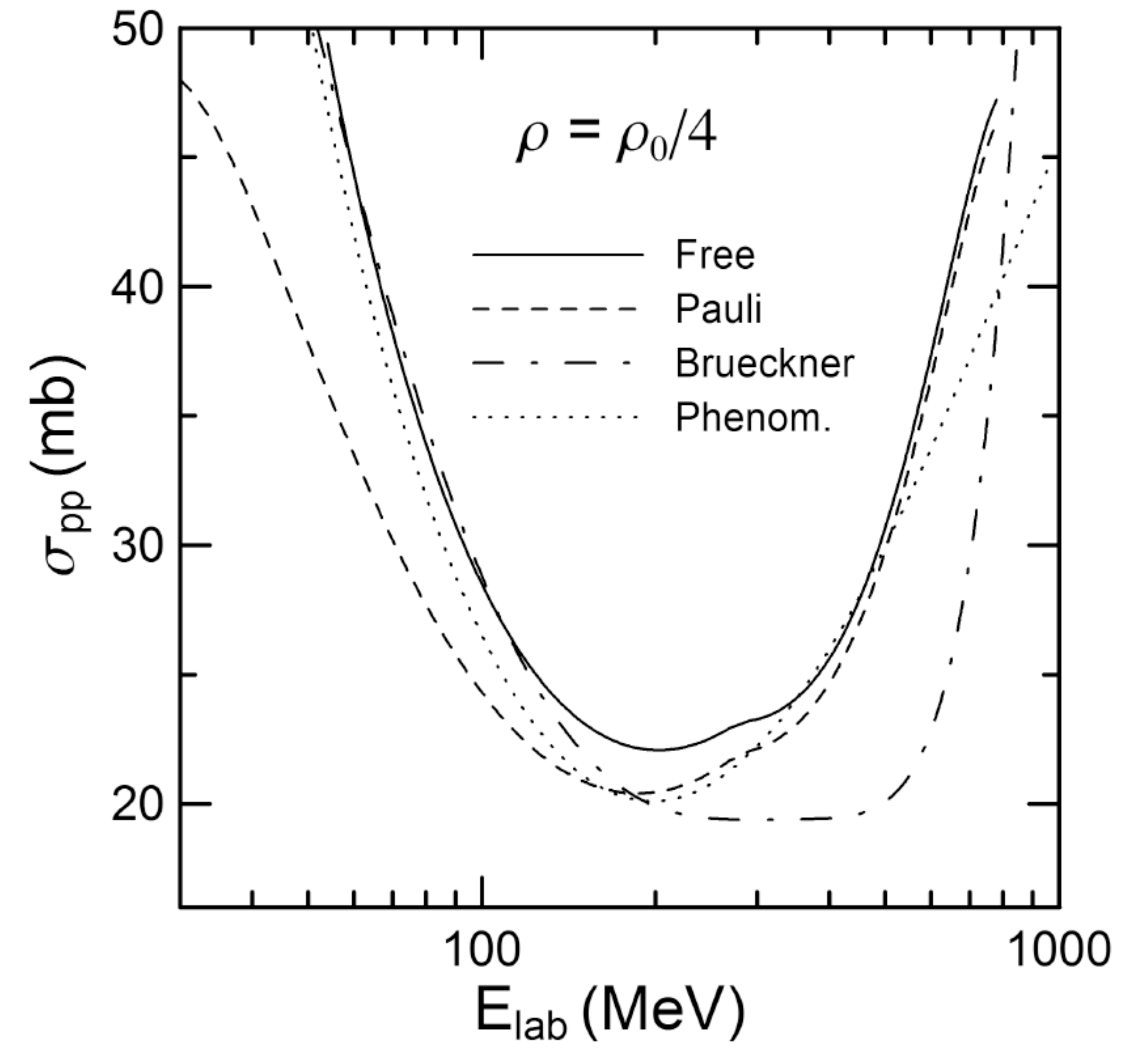}
\vspace{5mm} \caption{Same as in figure \ref{signpe}, but for pp collisions.} \label{sigppe}
\end{figure}

A usual approximation for the Pauli blocking is to assume that the effect of the $Q$ operator is equivalent to
a restricted angular integration in the domain (for symmetric nuclear matter)
\begin{equation}\label{eq:theta}
\frac{k_F^2-P^2-k^2}{2Pk} \leq  \cos \theta \leq
\frac{P^2+k^2-k_F^2}{2Pk}. 
\end{equation}
The integral in Eq.~(\ref{snngeo})
 becomes zero if the upper limit is negative (as determined
by the condition in Eq.~(\ref{12})), whereas the full
integration range is used if the upper limit is greater than 1.
(Notice that the average angle $\theta$ in Eq.~(\ref{eq:theta}),
namely the angle between the directions of ${\bf k}$ and ${\bf P}$,
is also the colatitude of ${\bf k}$ in a coordinate system where the
z-axis is along ${\bf P}$ and, thus, in such a reference frame it
coincides with the scattering angle to be integrated over in
Eq.~(\ref{snngeo})). The method of using
Eqs. \eqref{snngeo} and \eqref{eq:theta}, is not correct and misses an important part of the Pauli blocking geometry, as we show next.

A geometric description of the Pauli operator $Q$ was first studied by Clementel and Villi \cite{CV55} who
obtained an analytical expression for the scattering of a nucleon on a nucleon Fermi gas.
 By using the local density approximation, their work have been widely used to describe
Pauli-blocking in nucleon-nucleus scattering.  Much later, in Ref. \cite{Ber86} (see also Appendix C of Ref. \cite{HRB91}), an expression was obtained for the geometrical $Q$ operator for nucleon scattering in  asymmetric nuclear matter, involving two Fermi momentum spheres, one for the proton and another for the neutron. In contrast to Eq. \eqref{snngeo},  the expression obtained in Ref. \cite{Ber86} allows for NN-scattering with the relative momentum vector lying outside the symmetry axis of the two Fermi gas system.

As shown in Refs. \cite{Ber86,HRB91}, the  Pauli blocking projection yields an average nucleon-nucleon cross section for two Fermi gases with relative momenta ${\bf k}_0$ (see figure 30 of Ref. \cite{HRB91}) given by  
\begin{align}
\sigma_{NN}(k,\rho_1,\rho_2) &  =\int{\frac{d^{3}k_{1}d^{3}k_{2}}{(4\pi
k_{F1}^{3}/3)(4\pi k_{F2}^{3}/3)}}\ \nonumber\\
&  \times{\frac{2q}{k_0}}\ \sigma_{NN}^{free}(q)\
{\frac{\Omega_{Pauli}}{4\pi
}}\ , \label{ave}%
\end{align}
where  $2\mathbf{q}%
=\mathbf{k}_{1}-\mathbf{k}_{2}-\mathbf{k}_0$.  

Pauli-blocking enters through the restriction that the magnitude of the final nucleon momenta,
$|\mathbf{k^{\prime}}_{1}|$ and $|\mathbf{k^{\prime}}_{2}|$, lie outside the
Fermi spheres, with radii, $k_{F1}$ and $k_{F2}$. This leads to a limited fraction of the solid angle into which the nucleons can scatter, $\Omega_{Pauli}$.
It reads \cite{Ber86}
\begin{equation}
\Omega_{Pauli}=4\pi-2(\Omega_{a}+\Omega_{b}-{\bar{\Omega}})\ , \label{pauli}%
\end{equation}
where $\Omega_{a}$ and $\Omega_{b}$ specify the excluded solid angles for each
nucleon, and $\bar{\Omega}$ represents the geometric intersection of the solid angles $\Omega_{a}$
and $\Omega_{b}$,
\begin{equation}
\Omega_{a}=2\pi(1-\cos\theta_{a})\ ,\ \ \ \ \ \ \ \Omega_{b}=2\pi(1-\cos
\theta_{b})\ , \label{Omega}%
\end{equation}
where
\begin{align}
\cos\theta_{a}  &  =(p^{2}+q^{2}-k_{F1}^{2})/2pq\ ,\ \ \ \ \ \ \ \nonumber\\
\cos\theta_{b}  &  =(p^{2}+q^{2}-k_{F2}^{2})/2pq\ , \label{thetaab}%
\end{align}
with $2\mathbf{p}%
=\mathbf{k}_{1}+\mathbf{k}_{2}+\mathbf{k}_0$. 
\begin{figure}[!t]
\centering
\includegraphics[totalheight=8.0cm]{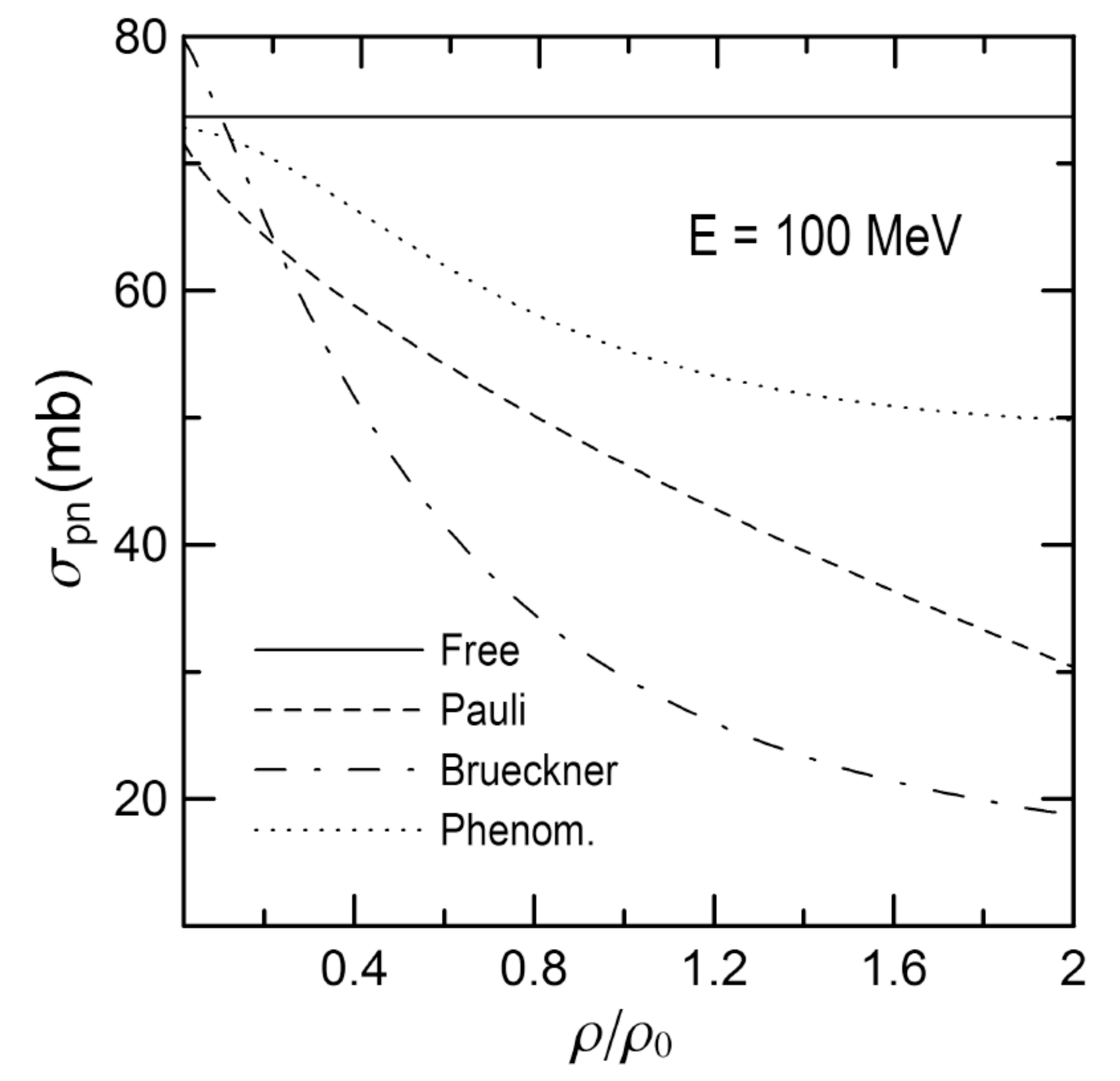}
\vspace{5mm} 
\caption{Parametrizations of proton-neutron cross sections as a function of the nuclear matter density (in units of $\rho_0=0.17$ fm$^{-3}$). The solid line is the parametrizarion  of the free $\sigma_{pn}$ cross section given by Eq. \eqref{signn2}.  The other curves include medium effects for symmetric nuclear matter for laboratory energy $E=100$ MeV. The dashed curve includes the geometrical effects of Pauli blocking, as described by Eq.  \eqref{VM1}. The dashed-dotted curve is the result of the Brueckner theory, Eq. \eqref{brueckner}, and the dotted curve is the phenomenological parametrization, Eq. \eqref{pheno}.} \label{sigpnr}
\end{figure}

For $\bar{\Omega}$  there are two possibilities:
\begin{align}
(1)\ \ \ \ \ {\bar{\Omega}} &  =\Omega_{i}(\theta,\theta_{a},\theta
_{b})+\Omega_{i}(\pi-\theta,\theta_{a},\theta_{b})\ ,\ \ \ \ \nonumber\\
&  \mathrm{for}\ \ \theta+\theta_{a}+\theta_{b}>\pi\nonumber\\
(2)\ \ \ \ \ {\bar{\Omega}} &  =\Omega_{i}(\theta,\theta_{a},\theta
_{b})\ ,\ \ \ \ \mathrm{for}\ \ \theta+\theta_{a}+\theta_{b}\leq
\pi\ ,\label{O1}%
\end{align}
\newline where $\theta$ is given by
\begin{equation}
\cos\theta=(k^{2}-p^{2}-b^{2})/2pb,\label{O2}%
\end{equation}
where $\mathbf{b}%
=\mathbf{k}-\mathbf{p}$.

The solid angle $\Omega_{i}$ has the following values
\begin{align}
(a)\ \ \ \ \ \Omega_{i} &  =0\ ,\ \ \ \mathrm{for}\ \ \theta\geq\theta
_{a}+\theta_{b}\nonumber\\
(b)\ \ \ \ \ \Omega_{i} &  =2\left[  \cos^{-1}\left(  \gamma_{ab}\right)
+\cos^{-1}\left(  \gamma_{ba}\right)  \right.  \nonumber\\
&  -\cos\theta_{a}\cos^{-1}\left(  {\frac{\cos\theta_{b}-\cos\theta\cos
\theta_{a}}{\sin\theta\sin\theta_{a}}}\right)  \nonumber\\
&  -\left.  \cos\theta_{b}\cos^{-1}\left(  {\frac{\cos\theta_{a}-\cos
\theta\cos\theta_{b}}{\sin\theta\sin\theta_{b}}}\right)  \right]  \nonumber\\
&  \mathrm{for}\ \ \ |\theta_{b}-\theta_{a}|\leq\theta\leq\theta_{a}%
+\theta_{b}\ ,\nonumber\\
(c)\ \ \ \ \ \Omega_{i} &  =\Omega_{b}\ \ \ \mathrm{for}\ \ \ \theta_{b}%
\leq\theta_{a},\ \theta\leq|\theta_{b}-\theta_{a}|\ ,\nonumber\\
(d)\ \ \ \ \ \Omega_{i} &  =\Omega_{a}\ \ \ \mathrm{for}\ \ \ \theta_{a}%
\leq\theta_{b},\ \theta\leq|\theta_{b}-\theta_{a}|\ ,\label{big}%
\end{align}
where
\[
\gamma_{jm}={\frac{\cos\theta_{m}-\cos\theta\cos\theta_{j}}{\sin\theta
_{j}(\cos^{2}\theta_{j}+\cos^{2}\theta_{m}-2\cos\theta\cos\theta_{j}\cos
\theta_{m})^{1/2}}.}%
\]

The integrals over $\mathbf{k}_{1}$ and $\mathbf{k}_{2}$ in
(\ref{ave}) reduce to a five-fold integral due to cylindrical
symmetry. The remaining integrals have to be performed numerically. 
One sees that for two Fermi gases the problem is much more complicated than the one studied in Ref. \cite{CV55}.
For symmetric nuclear matter, i.e. $k_F\equiv k_{F1}=k_{F2}$ the problem is still much more complicated than implied by Eq. \eqref{snngeo}, 
although many of the terms above simplify because in this case $\theta_a=\theta_b$ \cite{Ber01}. 

The numerical calculations can be simplified if we assume that  the free nucleon-nucleon cross section entering Eq. \eqref{ave} is isotropic.
 This is another rough approximation because the anisotropy of the free NN cross section is markedly manifest at large energies  \cite{LM:1993,LM:1994}.
 In the isotropic case, we  have devised a formula which fits the numerical integration in Eq. \eqref{ave} to within 1\%. The parametrization reads 
\begin{eqnarray}
\sigma_{NN}(E,\rho_1,\rho_2) &=&\sigma_{NN}^{free}(E){1 \over 1+1.892\left({\displaystyle{|\rho_1-\rho_2|\over \tilde{\rho}\rho_0}}\right)^{2.75}}\nonumber \\
&\times& 
\left\{
\begin{array}
[c]{c}%
\displaystyle{1-{37.02 \tilde{\rho}^{2/3}\over E}}, \ \ \   {\rm if} \ \ E>46.27 \tilde{\rho}^{2/3}\\ \, \\
\displaystyle{{E\over 231.38\tilde{\rho}^{2/3}}},\ \ \ \ \  {\rm if} \ \ E\le 46.27 \tilde{\rho}^{2/3}\end{array}
\right.
\label{VM1}
\end{eqnarray}
where $E$ is the laboratory energy in MeV, $\tilde{\rho}=(\rho_1+\rho_2)/\rho_0$,  with $\rho_0=0.17$ fm$^{-3}$.

\begin{figure}[!t]
\centering
\includegraphics[totalheight=8.0cm]{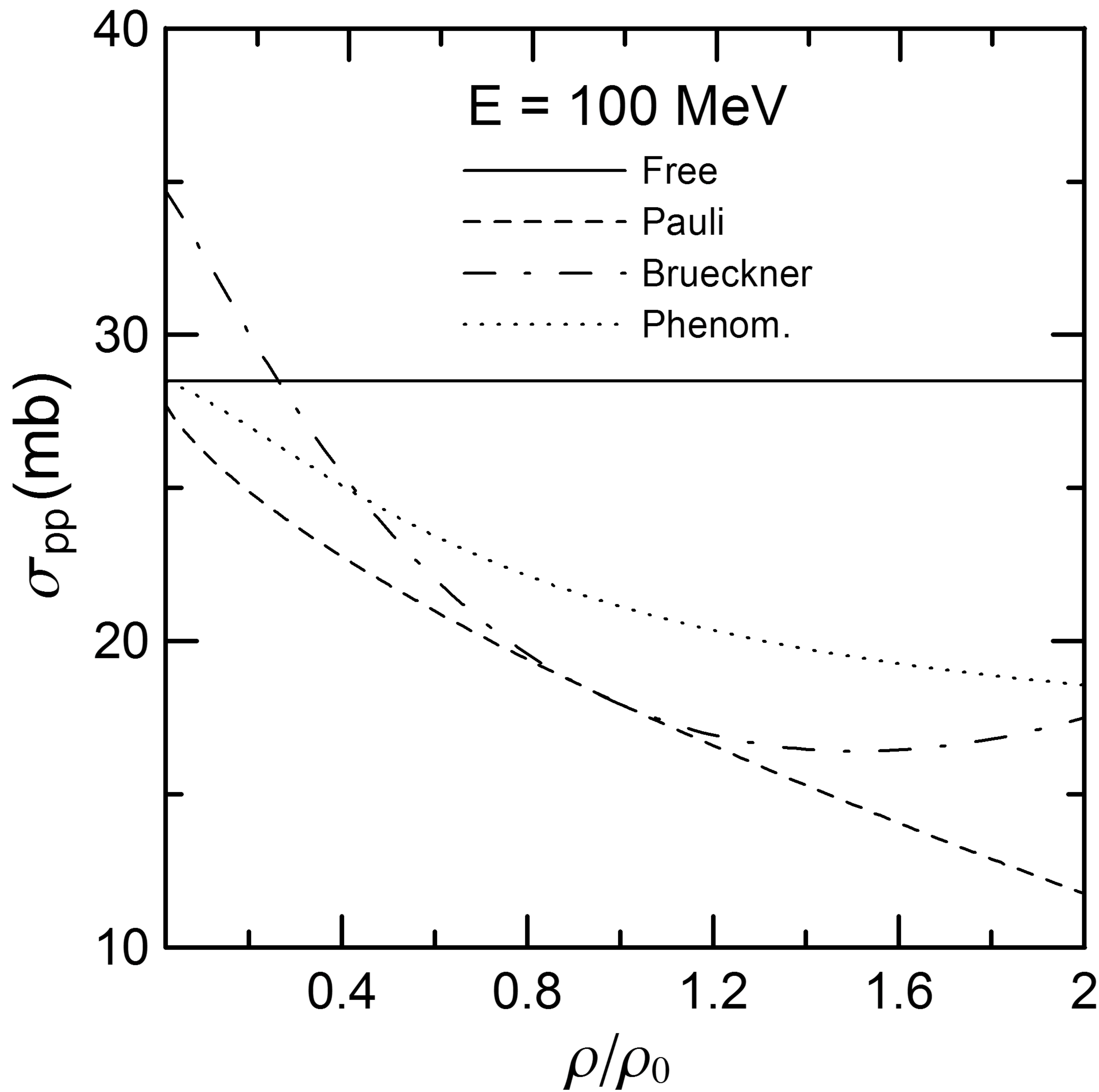}
\vspace{5mm} 
\caption{Same as in figure \ref{sigpnr}, but for proton-proton collisions.} \label{sigppr}
\end{figure}

The Brueckner method goes beyond a treatment of Pauli blocking, generating medium effects from  nucleon-nucleon potentials  such as the Bonn potential. An example is the work presented in Ref. \cite{LM:1993,LM:1994}, where a simple parametrization was given, which we will from now on refer as Brueckner theory. It reads (the misprinted factor 0.0256 in Ref. \cite{LM:1994} has been corrected to 0.00256)
\begin{align}
\sigma_{np}  &  = \left[ 31.5 +0.092\left| 20.2-E^{0.53}\right|^{2.9}\right] {1+0.0034E^{1.51} \rho^2\over 1+21.55\rho^{1.34}} \nonumber\\
\sigma_{pp}  &  = \left[ 23.5 +0.00256\left( 18.2-E^{0.5}\right)^{4.0}\right] {1+0.1667E^{1.05} \rho^3\over 1+9.704\rho^{1.2}}   \label{brueckner}
\end{align}

A modification of the above parametrization was done in Ref. \cite{Xian98}, which consisted in combining the free nucleon nucleon cross sections parametrized in Ref. \cite{Cha90} with the Brueckner theory results of Ref. \cite{LM:1993,LM:1994}. Their parametrization, which tends to reproduce better the nucleus-nucleus reactions cross sections, is 
\begin{align}
\sigma_{np}  &  = \left[ -70.67-18.18\beta^{-1}+25.26\beta^{-2}+113.85\beta\right] \nonumber \\
&\times {1+20.88E^{0.04} \rho^{2.02}\over 1+35.86\rho^{1.9}} \nonumber\\
\sigma_{pp}  &  = \left[ 13.73-15.04\beta^{-1}+8.76\beta^{-2}+68.67\beta^{4}\right]\nonumber\\
&\times {1+7.772E^{0.06} \rho^{1.48}\over 1+18.01\rho^{1.46}},   \label{pheno}
\end{align}
where $\beta=\sqrt{1-1/\gamma^2}$ and $\gamma=E{\rm[MeV]}/931.5+1$.
We will denote Eq. \eqref{pheno} as the phenomenological parametrization.

In figures (\ref{signpe}-\ref{sigppr}) we compare the several parametrizations above and we postpone the discussion of their details to section III.

\subsection{Nucleon knockout reactions}

The momentum distributions of the projectile-like residues in one-nucleon
knockout are a measure of the spatial extent of the wavefunction of the struck
nucleon, while the cross section for the nucleon removal scales with the
occupation amplitude, or probability (spectroscopic
factor), for the given single-particle configuration in the projectile ground
state. The longitudinal momentum distributions are given by (see, e.g.,
Refs. \cite{HBE96,BH04,BG06}) 
\begin{align}
\frac{d\sigma_{\mathrm{str}}}{dk_{z}}  &  = (C^2S) \frac{1}{2\pi%
}\frac{1}{2l+1}\sum_{m}\int_{0}^{\infty}d^{2}b_{n} \left[  1-\left\vert
S_{n}\left(  b_{n}\right)  \right\vert ^{2}\right]  \nonumber\\
&  \times  \int_{0}^{\infty}%
d^{2}\rho\ \left\vert S_{c}\left(  b_{c}\right)  \right\vert ^{2}\left\vert \int_{-\infty}^{\infty}dz \exp\left[  -ik_{z}z\right]
\psi_{lm}\left(  \mathbf{r}\right)  \right\vert ^{2},\label{strL}%
\end{align}
where $k_{z}$ represents the longitudinal component of $\mathbf{k}_{c}$ (final momentum of the
core of the projectile nucleus)
and $(C^2S)$ is the spectroscopic factor, and $\psi_{lm}\left(  \mathbf{r}\right)$
is the wavefunction of the core plus (valence) nucleon system $(c+n)$ in a state
with single-particle angular momentum $l,m$.  In this equation,
$\mathbf{r\equiv}\left(\rho,z,\phi\right)  =\mathbf{r}_{n}-\mathbf{r}_{c}$, so that
\begin{eqnarray}
b_{c}    &=\left\vert \mathbf{\mbox{\boldmath$\rho$}}-\mathbf{b}
_{n}\right\vert =\sqrt{\rho^{2}+b_{n}^{2}-2\rho\ b_{n}\cos\phi  }\nonumber \\
  &=\sqrt{r^{2}\sin^{2}\theta+b_{n}^{2}-2r\sin\theta\ b_{n}\cos
\phi  }.
\end{eqnarray}

\begin{figure}[!t]
\centering
\includegraphics[totalheight=8.0cm]{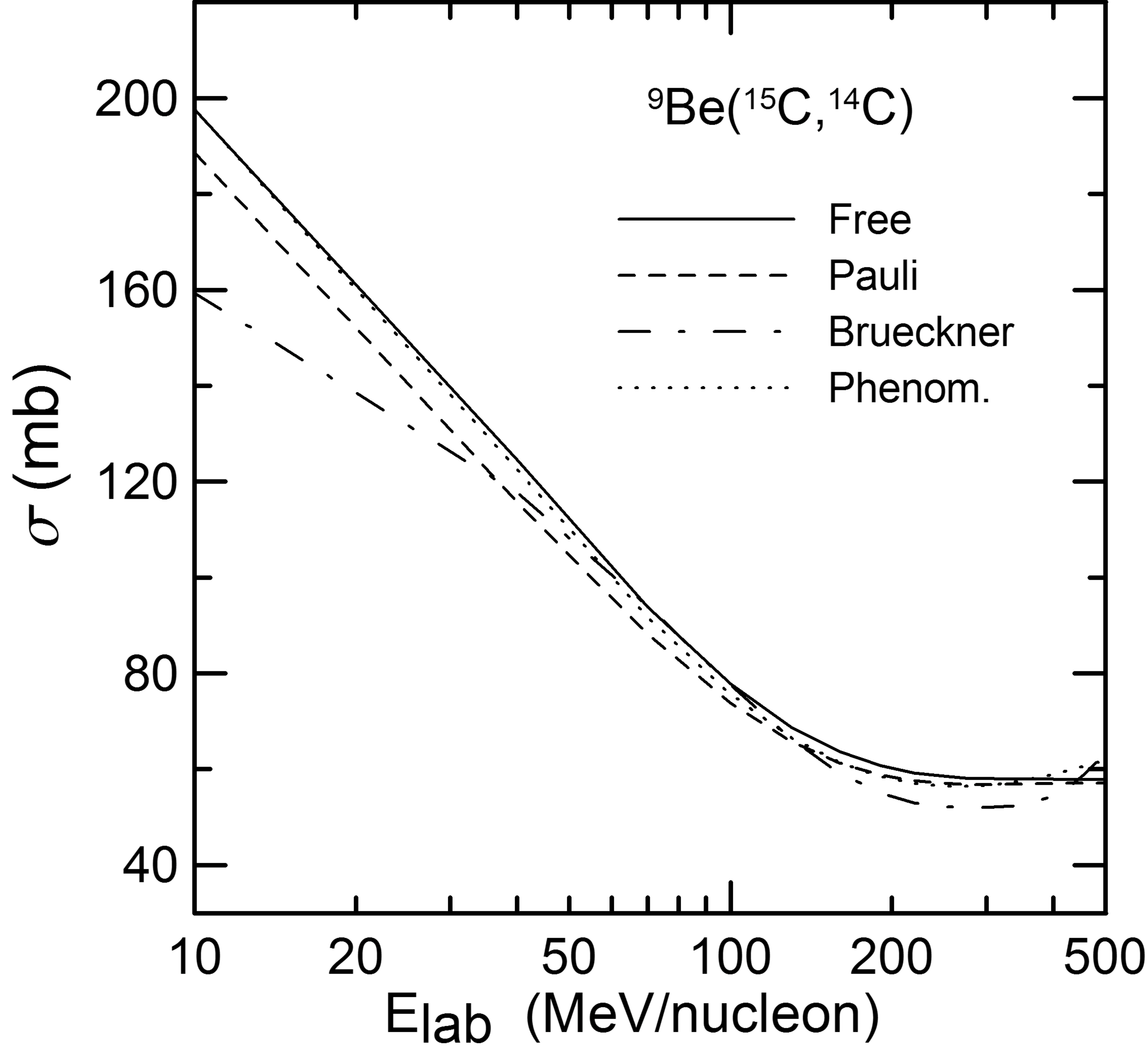}
\vspace{5mm} 
\caption{Total  knockout cross sections for removing the $l=0$ halo neutron of $^{15}$C, bound by 1.218 MeV, in the reaction $^9$Be($^{15}$C,$^{14}$C$_{gs}$). The solid curve is obtained with the use of free nucleon-nucleon cross sections.  The dashed curve includes the geometrical effects of Pauli blocking, as described by Eq.  \eqref{VM1}. The dashed-dotted curve is the result using the Brueckner theory, Eq. \eqref{brueckner} and the dotted curve is the phenomenological parametrization, Eq. \eqref{pheno}.} \label{sko15c}
\end{figure}

$S_i(b)$ are the $S$-matrices for core-target and nucleon-target scattering
obtained from the nuclear ground-state 
densities and the nucleon-nucleon cross sections by the relation \cite{BD04}
$
S(b)=\exp\left[  i\phi(b)\right]  
$, with
\begin{equation} 
\phi
_{N}(b)=\frac{\sigma_{NN}}{4\pi}\int_{0}^{\infty}dq\ q\ \rho_{p}\left(  q\right)
\rho_{t}\left(  q\right)    J_{0}\left(  qb\right)
\ ,\label{eikphase}%
\end{equation}
where $\rho_{p,t}\left(  q\right)  $ is the Fourier transform of the nuclear
density of the projectile (nucleon or core) and the target nucleus, and
$\sigma_{NN}$ is the
nucleon-nucleon total cross section.
One needs to add the Coulomb phase to the nuclear eikonal phase of Eq. \eqref{eikphase}. This is done by using a 
sharp-cutoff expression for the Coulomb phase, as explained in Refs.  \cite{BH04,BG06}. 

The first term inside the integrals in Eq. \eqref{strL}, $1-|S_n|^2$, represents the
probability for the knockout of the nucleon from its location at $b_n$, whereas
the second integral carries the term $|S_c|^2$ which is the probability of
core survival at impact parameter $b_c$.  These results arise naturally by using
eikonal scattering waves \cite{BD04}. 
For the transverse momentum distributions, the same formalism yields
\begin{align}
\frac{d\sigma_{\mathrm{str}}}{d^{2}k^{\bot}_c}  &  =(C^2S)\frac{1}{(2\pi)^2}\frac{1}%
{2l+1} \int_{0}^{\infty}d^{2}b_{n} \left[  1-\left\vert S_{n}\left(
b_{n}\right)  \right\vert ^{2}\right] \nonumber\\
&  \times\sum_{m} \int_{-\infty}^{\infty}dz \left\vert \int d^{2}%
\rho \exp\left(  -i\mathbf{k}_{c}^{\perp}\mathbf{.\mbox{\boldmath$\rho$}}%
\right)  S_{c}\left(  b_{c}\right)  \psi_{lm}\left(  \mathbf{r}\right)
\right\vert ^{2},\label{strT}%
\end{align} 
where ${\bf k}_c^\bot$ is the perpendicular component of ${\bf k}_c$. 

\begin{figure}[!t]
\centering
\includegraphics[totalheight=8.0cm]{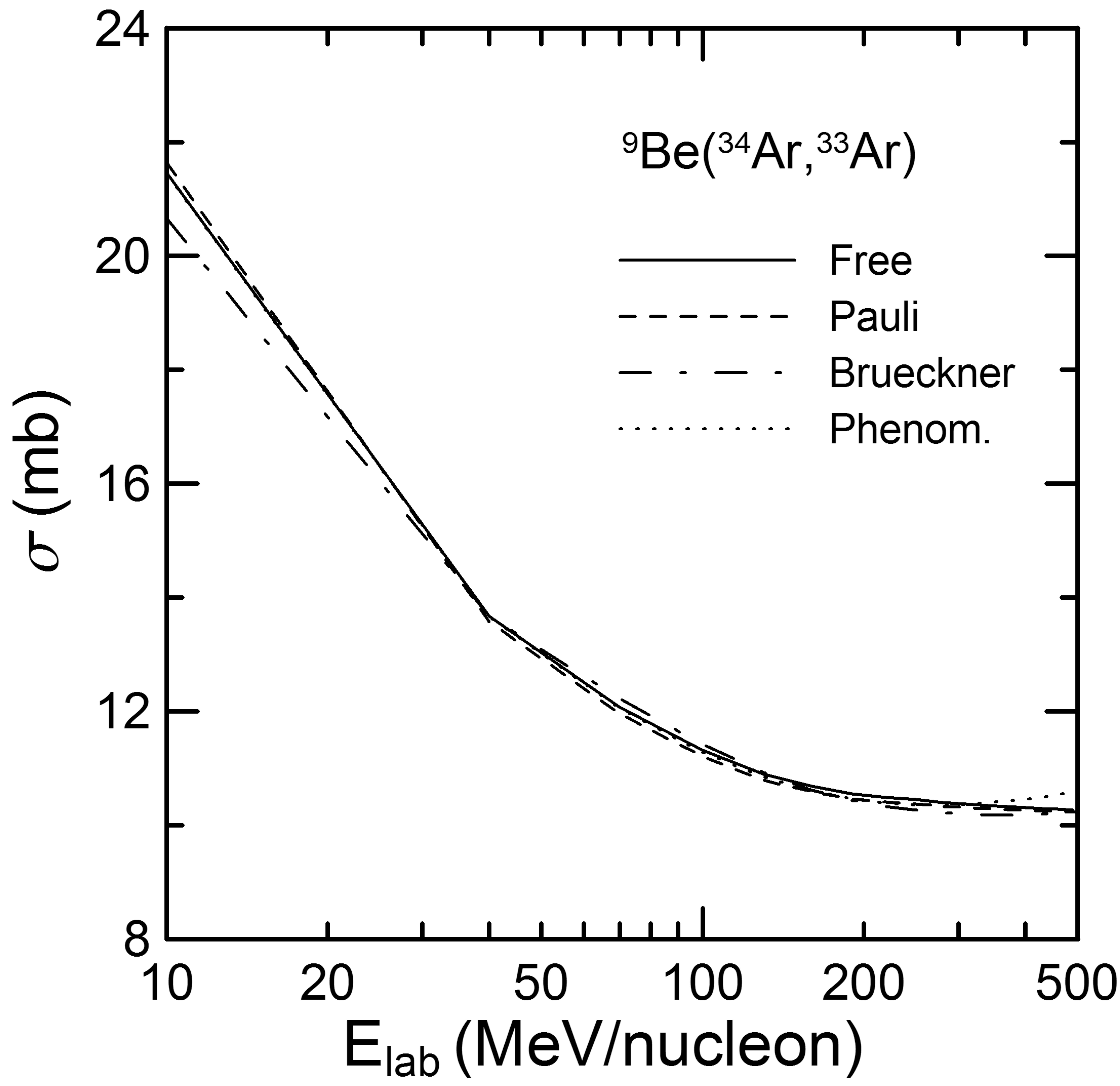}
\vspace{5mm} 
\caption{Same as in figure \ref{sko15c}, but for the removal of the $l=0$ neutron bound by  17.06 MeV in the reaction $^9$Be($^{34}$Ar,$^{33}$Ar(1/2$^+$)).} \label{sko34ar}
\end{figure}   

The total stripping cross section can be obtained by integrating
either Eq. (\ref{strL}) or Eq. (\ref{strT}). One obtains%
\begin{eqnarray}
\sigma_{\mathrm{str}}  &  =&(C^2S) \frac{2\pi}{2l+1}\ \sum _{m}\int_{0}^{\infty}db_{n}%
\ b_{n}\ \left[  1-\left\vert S_{n}\left(  b_{n}\right)  \right\vert
^{2}\right]  \ \nonumber\\
&  \times&\int d^{3}r\ \left\vert S_{c}\left(
b_c\right)  \right\vert ^{2}\left\vert
\psi_{lm}\left(  \mathbf{r}\right)  \right\vert ^{2}.
\end{eqnarray}

The total diffraction dissociation cross section is given by \cite{BG06}
\begin{eqnarray}
\sigma_{\mathrm{dif}}    &=&(C^2S) \frac{2\pi}{2l+1}\sum _{m} \int_{0}^{\infty}db_{n}%
\ b_{n}  \nonumber\\
&  \times&\left\{ \int d^{3}r   \bigg\vert S_{n}\left(  b_{n}\right)S_{c}\left(
b_c\right) \psi_{lm}\left(  \mathbf{r}\right)\bigg\vert
^{2}\right.\nonumber\\
&-& \sum_{m'}\left. \left\vert \int d^3r  \psi_{lm'}\left(  \mathbf{r}\right)  S_{c}\left(
b_c\right)  S_n(b_n) 
\psi_{lm}\left(  \mathbf{r}\right)  \right\vert^{2} \right\}.
\label{sigdiff}
\end{eqnarray}

To render the calculations practical,  for a nucleus-nucleus collision with a given impact parameter $b$, we have obtained an effective local density for protons and neutrons by taking the point along the impact parameter direction  where the two densities (one from the projectile and the other from the target) cross each other. This effective density
was then used in Eqs. \eqref{VM1},  \eqref{brueckner}, and \eqref{pheno}

In the following we will use a modified version of the code MOMDIS \cite{BG06} which includes the new aspects of momentum distributions discussed in this article. As we want to make a theoretical study of the medium effects of the nucleon-nucleon cross sections, we do not compare directly to experiments and we use spectroscopic factors $(C^2S)$ equal to the unity. To generate the wavefunctions and S-matrices, we use the same parameters as in Refs. \cite{BH04} and \cite{BG06}.

\section{Results and discussion}

There are marked differences between the parametrization of the Brueckner \eqref{brueckner}, the geometrical Pauli blocking \eqref{VM1} and the phenomenological one \eqref{pheno}. An example is given in figure \ref{signpe} where the several parametrizations of proton-neutron cross sections are shown as a function of the laboratory energy. The solid line is the parametrizarion  of the free $\sigma_{pn}$ cross section given by Eq. \eqref{signn2}.  The other curves include medium effects for symmetric nuclear matter for $\rho=\rho_0/4$, where $\rho_0=0.17$ fm$^{-3}$. The dashed curve includes the geometrical effects of Pauli blocking, as described by Eq.  \eqref{VM1}. The dashed-dotted curve is the result of the Brueckner theory, Eq. \eqref{brueckner}, and the dotted curve is the phenomenological parametrization, Eq. \eqref{pheno}.  The large deviation of the parametrization of the Brueckner results at large energies is not physical because Eq. \eqref{brueckner} is  only a good parametrization of the  Brueckner theory in the energy range of 50-300 \cite{LM:1993,LM:1994}. At energies above 300 MeV inelastic channels have to be incorporated. However, the other differences are real, especially those at lower energies. Pauli-blocking effectively reduces the in-medium  np cross section.  This is not so apparent in the phenomenological parametrization.

\begin{table}[ptb]%
\begin{tabular}
[c]{cccccc}\hline\hline
Reaction                             & $\sigma$       & Free & Pauli & Brueckner & Pheno. \\
\hline
$^9$Be($^{11}$Be,$^{10}$Be)          & $\sigma_{dif}$ & 47.6 & 36.9  & 45.7      & 45.2   \\
                                     & $\sigma_{str}$ & 151. & 144.  & 139.      & 149.   \\
                                     & $\sigma_{tot}$ & 198. & 181.  & 185.      & 194.   \\
\hline
$^9$Be($^{15}$C,$^{14}$C)            & $\sigma_{dif}$ & 25.3 & 19.9  & 21.3      & 24.0   \\
                                     & $\sigma_{str}$ & 99.8 & 95.8  & 96.5      & 98.5   \\
                                     & $\sigma_{tot}$ & 125. & 116.  & 118.      & 123.   \\
\hline
$^9$Be($^{34}$Ar,$^{33}$Ar($1/2^+$)) & $\sigma_{dif}$ & 2.69 & 2.63  & 2.66      & 2.68   \\
                                     & $\sigma_{str}$ & 11.0 & 10.9  & 11.0      & 11.0   \\
                                     & $\sigma_{tot}$ & 13.6 & 13.5  & 13.6      & 13.6   \\
\hline\hline
\end{tabular}\caption{Cross sections in mb at 40 MeV/nucleon for nucleon knockout of a few selected reactions.}%
\label{t4}
\end{table}

The observations above cannot be extended to the pp cross sections, which are shown in figure \ref{sigppe}. Here we see that the Pauli-blocking correction decreases the cross section  much more than in the other cases. Some important differences are also clearly visible at larger energies, $E\gtrsim 100$ MeV/nucleon.

Figure \ref{sigpnr} shows the  proton-neutron cross sections as a function of the nuclear matter density  (in units of $\rho_0=0.17$ fm$^{-3}$), for a proton laboratory energy of $E_{lab}=100$ MeV. The solid line is the parametrizarion  of the free $\sigma_{pn}$ cross section given by Eq. \eqref{signn2}.  The other curves include medium effects for symmetric nuclear matter. The dashed curve includes the geometrical effects of Pauli blocking, as described by Eq.  \eqref{VM1}. The dashed-dotted curve is the result of the Brueckner theory, Eq. \eqref{brueckner}, and the dotted curve is the phenomenological parametrization, Eq. \eqref{pheno}. Figure \ref{sigppr} shows the same as in figure \ref{sigpnr}, but for proton-proton collisions. One notices that the nucleon-nucleon cross sections differ appreciably at large densities but they become close to the free cross sections at low densities. 

To test the influence of the medium effects in nucleon knockout reactions, we consider the removal of the $l=0$ halo neutron of $^{15}$C, bound by 1.218 MeV,  and the $l=0$ neutron knockout from $^{34}$Ar, bound by 17.06 MeV. The reactions studied here are the $^9$Be($^{15}$C,$^{14}$C$_{gs}$)  and  $^9$Be($^{34}$Ar,$^{33}$Ar(1/2$^+$)). The total cross sections as a function of the bombarding energy are shown in figures \ref{sko15c} and \ref{sko34ar}. The solid curve is obtained with the use of free nucleon-nucleon cross sections.  The dashed curve includes the geometrical effects of Pauli blocking, as described by Eq.  \eqref{VM1}. The dashed-dotted curve is the result using the Brueckner theory, Eq. \eqref{brueckner}, and the dotted curve is the phenomenological parametrization, Eq. \eqref{pheno}.  

The medium effects due to different treatments are more visible  for the  $^9$Be($^{15}$C,$^{14}$C$_{gs}$) reaction. For  $^9$Be($^{34}$Ar,$^{33}$Ar(1/2$^+$)) the differences are almost not visible, as shown in figure \ref{sko34ar}. The same happens for the $l=2$ neutron removal reaction leading to a final $3/2^+$ level bound by 18.42 MeV. A similar behavior as for the  $^9$Be($^{15}$C,$^{14}$C$_{gs}$) reaction is found for the removal of the halo neutron in the nucleon knockout $^9$Be($^{11}$Be,$^{10}$Be) bound by 0.504 MeV. It is thus apparent that the corrections due to the medium effects are more evident for the knockout out from loosely bound states. Knockout reactions are also more sensitive to in-medium corrections of the nucleon-nucleon cross sections than the total reaction cross sections, as first pointed out in \cite{HRB91}. 

In table \ref{t4} we show our results for the stripping, diffraction dissociation and total nucleon cross section (in mb) for the knockout reactions  $^9$Be($^{11}$Be,$^{10}$Be), $^9$Be($^{15}$C,$^{14}$C), and $^9$Be($^{34}$Ar,$^{33}$Ar) at 40 MeV/nucleon. For the reaction $^9$Be($^{34}$Ar,$^{33}$Ar($3/2^+$)) the values are $\sigma_{dis}=2.36$ mb, $\sigma_{str}=9.16$ mb and $\sigma_{tot}=11.5$ mb and have the same value, within 1\%, for calculations using all of the NN cross section parametrizations, either Eq. (\ref{signn1}-\ref{signn2}), or  \eqref{VM1}, \eqref{brueckner} or \eqref{pheno}.

In table \ref{t5} we show the same  as in table \ref{t4} but for $E_{lab}=250$ MeV/nucleon.  For the reaction $^9$Be($^{34}$Ar,$^{33}$Ar($3/2^+$)) the values are $\sigma_{dis}=0.691$ mb, $\sigma_{str}=8.62$ mb and $\sigma_{tot}=9.32$ mb and have the same value, and as in table \ref{t4}, these values differ by less than 1\%, for all NN cross section parametrizations used in the calculations.

\begin{table}[ptb]%
\begin{tabular}
[c]{cccccc}\hline\hline
Reaction                             & $\sigma$       &   Free & Pauli & Brueckner & Pheno.\\
\hline
$^9$Be($^{11}$Be,$^{10}$Be)          & $\sigma_{dif}$ &   11.0 & 10.3  &  8.64     & 10.0  \\
                                     & $\sigma_{str}$ &   74.4 & 73.0  &  66.0     & 71.7  \\
                                     & $\sigma_{tot}$ &  85.8  & 83.1  & 75.0      & 81.7  \\
\hline
$^9$Be($^{15}$C,$^{14}$C)            & $\sigma_{dif}$ & 5.14   & 4.78  & 3.90      & 4.63  \\
                                     & $\sigma_{str}$ & 53.4   & 52.3  & 48.2      & 51.8  \\
                                     & $\sigma_{tot}$ & 58.5   & 57.1  & 52.1      & 56.4  \\
\hline                                     
$^9$Be($^{34}$Ar,$^{33}$Ar($1/2^+$)) & $\sigma_{dif}$ & 0.801  & 0.785 & 0.749     & 0.778 \\
                                     & $\sigma_{str}$ & 9.62   & 9.55  & 9.47      & 9.55  \\
                                     & $\sigma_{tot}$ & 10.5   &  10.4 &    10.2   & 10.4  \\
\hline\hline
\end{tabular}
\caption{Cross sections in mb at 250 MeV/nucleon for nucleon knockout of a few selected reactions.}%
\label{t5}
\end{table}

\begin{figure}[!t]
\centering
\includegraphics[totalheight=8.0cm]{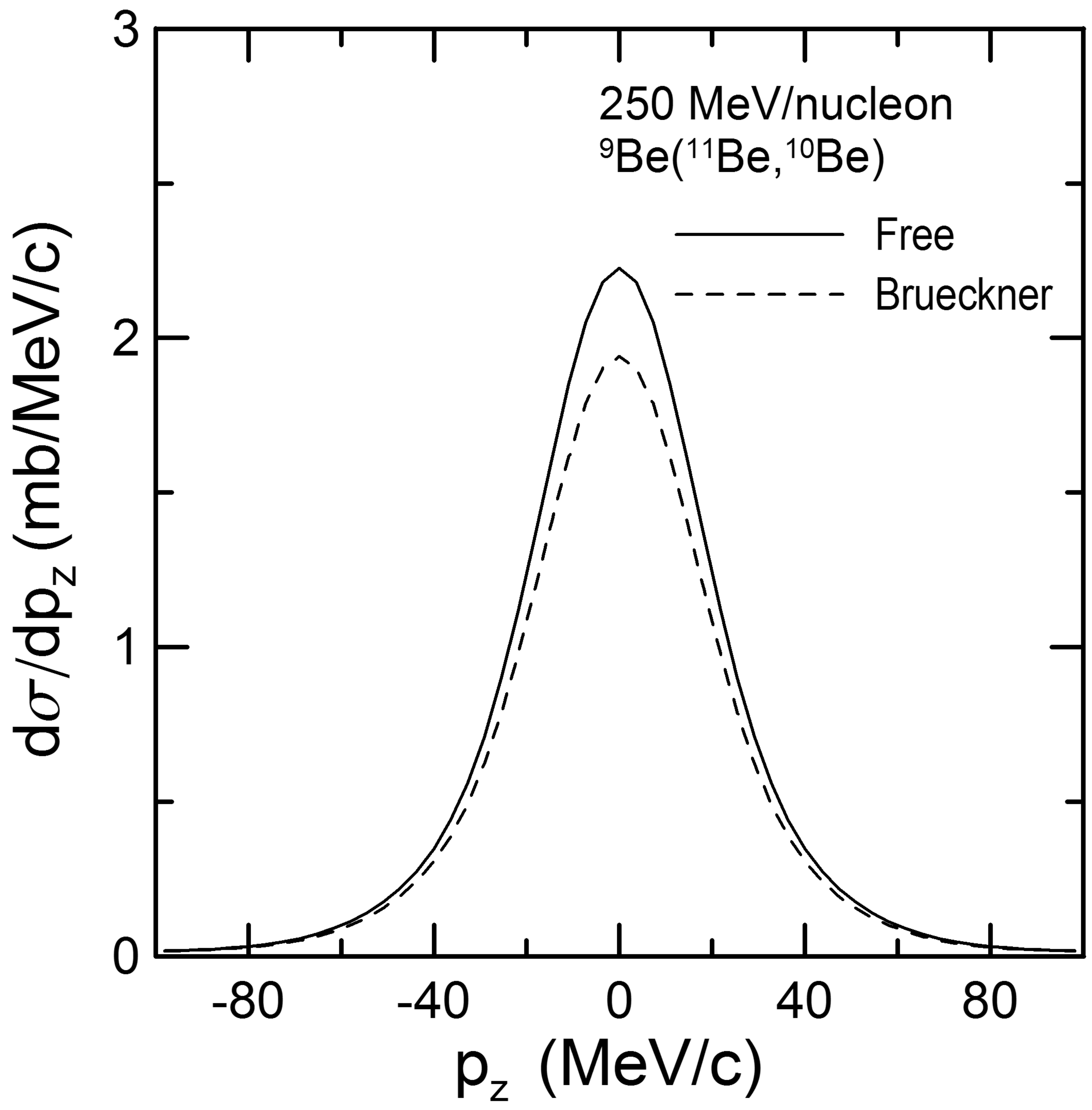}
\vspace{5mm} 
\caption{Longitudinal momentum distribution for the residue in
the $^9$Be($^{11}$Be,$^{10}$Be),  reaction at 250 MeV/nucleon. The dashed curve is the cross
section calculated using the NN cross section from the Brueckner theory, Eq. \eqref{brueckner} and the solid curve is obtained  the free cross section, Eq. \eqref{pheno}.} \label{dsdpz}
\end{figure}

In figure \ref{dsdpz}  we plot the longitudinal momentum distributions for the reaction $^9$Be($^{11}$Be,$^{10}$Be),  at 250 MeV/nucleon. The calculations are done using Eq. \eqref{strL}. The diffraction dissociation cross sections have been calculated using the same profile of the momentum distribution due to stripping, but with the total cross section normalized to Eq. \eqref{sigdiff}. The different contributions (stripping and diffraction dissociation) to this reaction are given in table \ref{t5}. In figure \ref{dsdpz}  the dashed curve is the cross section calculated using the NN cross section from the Brueckner theory, Eq. \eqref{brueckner} and the solid curve is obtained  the free cross section, Eq. \eqref{pheno}. One sees  that the momentum distributions are reduced by 10\%, about the same as  the total cross sections, but the shape remains basically unaltered. If one rescales the dashed curve to match the solid one, the differences in the width are not visible.
We do not show the momentum distributions using the other two (Pauli and phenomenological) NN cross sections as their shapes are the same as for the Brueckner case and only the area below the curve (total knockout cross section) changes.

\begin{figure}[!t]
\centering
\includegraphics[totalheight=8.0cm]{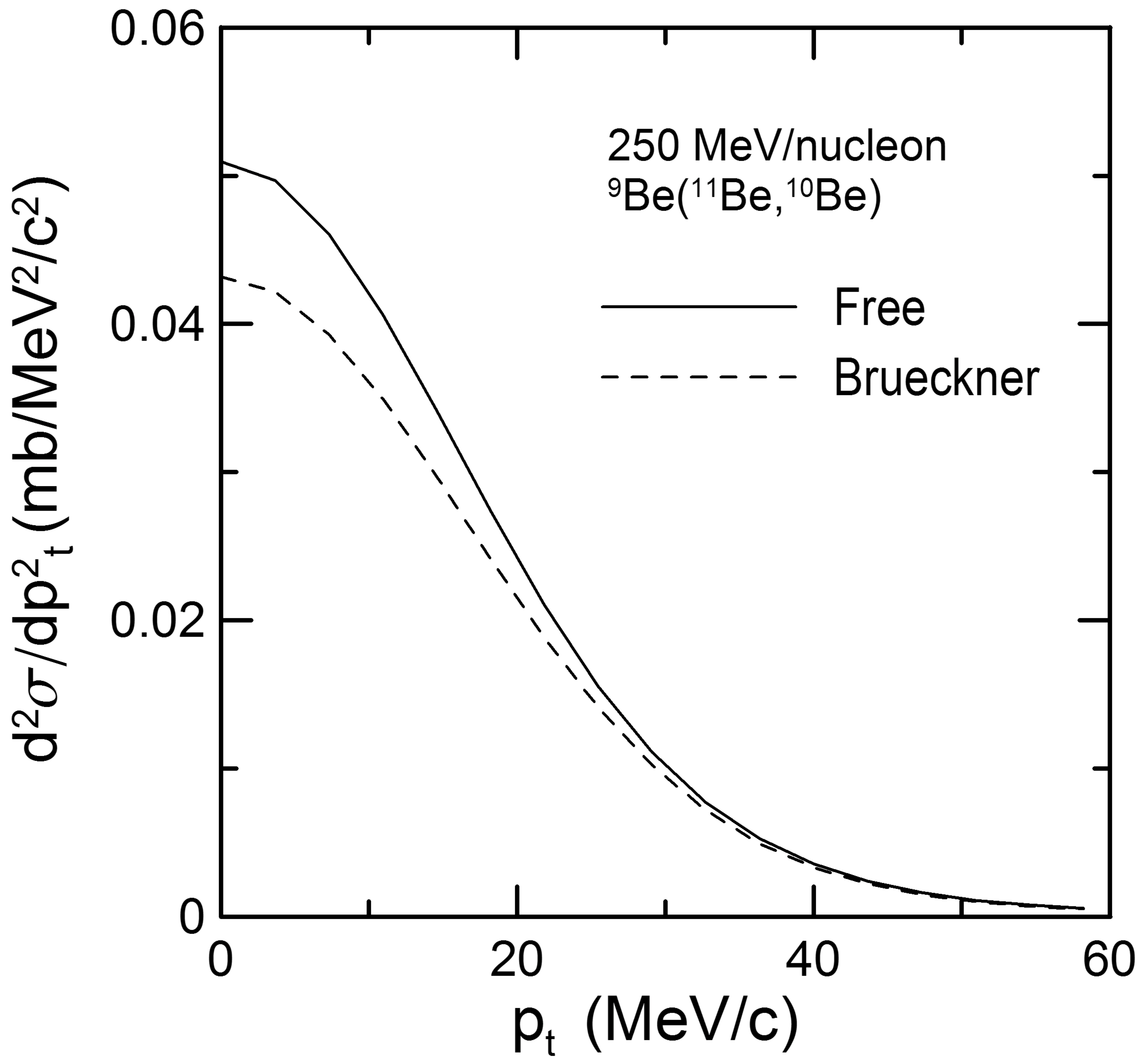}
\vspace{5mm} 
\caption{Same as in figure \ref{dsdpz}, but for the transverse momentum distribution.} \label{dsdt}
\end{figure}

In figure \ref{dsdt}  we plot the transverse momentum distributions for the reaction $^9$Be($^{11}$Be,$^{10}$Be),  at 250 MeV/nucleon. The calculations are done using Eq. \eqref{strT}. As with figure \ref{dsdpz},  the dashed curve in figure \ref{dsdt} is the cross section calculated using the NN cross section from the Brueckner theory, Eq. \eqref{brueckner}, and the solid curve is obtained  the free cross section, Eq. \eqref{pheno}. The changes on the profile of the momentum distribution are again visible, what is again ascribed to the difference of about 10\% between the total cross sections. The form of the momentum distributions are the same if the two curves are scaled to have the same area.
  
 These results clearly show that the effects of nucleon-nucleon scattering in the medium on knockout reactions are worth considering, specially for reactions involving loosely-bound halo nuclei.  It is not clear however which of the several parametrizations of  medium effects is more adequate for the precision required by experiments.  

\section{Summary}

The present work has extended the theory of one-nucleon
stripping and diffraction dissociation reactions  to cover the dependence of the nucleon knockout cross sections
and  momentum distributions on the medium modifications of the nucleon-nucleon cross sections.
We included the most commonly used parametrizations in the literature and compared the effects of Pauli blocking from a simple
geometrical picture to a more elaborated Dirac-Brueckner calculation, as well as phenomenological parametrizations.
We have shown that the density dependences vary rather strongly from model to model for reasons which are not yet clear.

We have also shown that the nucleon knockout reactions involving halo nuclei are more sensitive to medium modifications of the 
NN cross section than in the case of the removal of more bound nucleons. The changes amount to 10\% in some cases, especially at lower energies.
But due to the average of the nucleon-nucleon cross sections over the local densities, the changes are not always predictable at higher energies. 
The stripping and diffraction dissociation cross sections decrease and increase in the same way whenever the NN cross sections decrease or increase from one parametrization to another.
 
The momentum distributions, are not appreciably different, except for their absolute normalization, when the nucleon-nucleon cross sections change with medium modifications. This has been verified for both longitudinal and transverse momentum distributions.

The simple study of the density-dependent NN cross-section adopted in this
work shows that the calculations are sometimes sensitive to the value of the density-dependence method under consideration.
Besides Pauli-blocking and medium changes in the NN cross section, Fermi-motion should probably play an important
role in the nucleon knockout reactions and is worth further investigation. 

No attempt has been done to compare to experimental results, which would probably affect the extracted values of the spectroscopic strengths in reactions with rare isotopes. This certainly deserves further theoretical studies. 

It is also worth mentioning that the magnitude of the corrections observed
in this work, of about 10\% for the total cross sections are based on the optical limit (OL) 
of the Glauber multiple scattering theory. The optical limit means that only single binary NN collisions are included. In the 
present work correlations within the projectile and target wave functions have been neglected. These have  been studied, e.g. in Ref. \cite{FV77}, or more recently in Ref. \cite{AS00}. In these references, the influence of these correlations on the calculations has been studied, and found to be also of the order of 10\%. This
is of the same order of magnitude as the corrections observed in the present work. It is not clear if these two unrelated corrections will add up to a larger correction of the knockout cross sections, which could in fact modify appreciably the spectroscopic factors published in the literature where such corrections where not included. This also qualifies for further investigation. 

It is worthwhile mentioning that medium modifications of nucleon-nucleon scattering have also been studied in several publications related to (p,2p) reactions (see, e.g. Refs \cite{JM73,SC86,Krei95}). The medium effects were shown to play an important role on the total cross sections and on the spin observables. 

\bigskip

\section*{Acknowledgements}

C.~B. thanks Dr. Lie-Wen Chen for useful conversations. C.~B. acknowledges support by the U.\thinspace S.\ Department of
Energy under grant No. DE-FG02-08ER41533, DE-FC02-07ER41457
(UNEDF, SciDAC-2) and the Research Corporation. C.D.C. acknowledges the Brazilian agency FAPESP (Funda\c c\~ao de Amparo a Pesquisa do Estado de S\~ao Paulo)  for financial support 
and the hospitality of the  Department of Physics and Astronomy of the Texas A\&M University-Commerce, where this work was done.


\begin{thebibliography}{30}

\bibitem{BM92}C.A. Bertulani and K.W. McVoy, Phys. Rev. C 46 (1992) 2638

\bibitem{Gregers}P.\ G.\ Hansen, Nature 334 (1998) 194.

\bibitem{Tostevin1999}J.\ A.\ Tostevin, J.\ Phys.\ G 25 (1999) 735.

\bibitem{han03}P.\,G.\ Hansen and J.\,A. Tostevin, Annu. Rev. Nucl. Part.
Sci. {53} (2003) 219.

\bibitem{Gade2008a}
A.~Gade, P.~Adrich, D.~Bazin, M.~D. Bowen, B.~A. Brown, C.~M. Campbell, J.~M.
  Cook, T.~Glasmacher, P.~G. Hansen, K.~Hosier, S.~McDaniel, D.~McGlinchery,
  A.~Obertelli, K.~Siwek, L.~A. Riley, J.~A. Tostevin, D.~Weisshaar,
  Phys.~Rev.~C 77 (2008) 044306.
  
\bibitem{BH04}C.A. Bertulani and P.G. Hansen, Phys. Rev. C 70, 034609
(2004).

\bibitem{AS00} B. Abu-Ibrahim and Y. Suzuki, Phys. Rev. C 61, 051601(R) (2000).

\bibitem{AOS03} B. Abu-Ibrahim, Y. Ogawa, Y. Suzuki, and I. Tanihata, Comp. Phys. Comm. 151, 369 (2003).

\bibitem{Tostevin2006}J.\ A.\ Tostevin and B.\ A.\ Brown, Phys.\ Rev.\ C 74
  (2006) 064604.

\bibitem{Sim09}E. C. Simpson, J. A. Tostevin, D. Bazin, B. A. Brown and
  A. Gade,  Phys. Rev. Lett. 102 (2009) 132502.

\bibitem{Simpson2009}E.\ C.\ Simpson, J.\ A.\ Tostevin, D.\ Bazin, and A.\
  Gade, Phys.\ Rev.\ C 79 (2009) 064621.

\bibitem{Li08}Bao-An Li, Lie-Wen Chen and Che Ming Ko, Physics Reports 464, 113 (2008).

\bibitem{LM:1993}G.~Q.~Li and R.~Machleidt, Phys. Rev. C 48, 1702
(1993).

\bibitem{LM:1994}G.~Q.~Li and R.~Machleidt, Phys. Rev. C 49, 566
(1994).

\bibitem{FK06}F. Sammarruca and P. Krastev, Phys. Rev. C 73, 014001 (2006).

\bibitem{Sam08}F. Sammarruca, Phys. Rev. C 77, 047301 (2008).

\bibitem{HRB91}M. Hussein, R. Rego and C.A. Bertulani, Physics Reports 201, 279 (1991).

\bibitem{pdgxnn}Particle Data Group. J. Phys. G 33, 1 (2006).

\bibitem{GWW58}L.C. Gomes, J.D. Walecka, and V.F. Weisskopf, Ann. Phys. (N.Y.) 3, 241 (1958).

\bibitem{CV55}E. Clementel and C. Villi, Nuovo Cimento 11 (1955) 176.

\bibitem{Ber86}C.A. Bertulani, Braz. J. Phys. 16, 380
(1986).

\bibitem{Ber01}C.A. Bertulani, J. Phys. G 27, L67 (2001).

\bibitem{Xian98}Cai Xiangzhou, Feng Jun, Shen Wenqing, Ma Yugang, Wang Jiansong, and Ye Wei, Phys. Rev. C58, 572 (1998).

\bibitem{Cha90}S. K. Charagi and S. K. Gupta, Phys. Rev. C 41, 1610 (1990).

\bibitem{HBE96}K. Hencken, G. Bertsch and H. Esbensen, Phys. Rev. C 54, 3043 (1996).

\bibitem{BG06}C.A. Bertulani and A. Gade, Comp. Phys. Comm. 175, 372 (2006). 

\bibitem{BD04}C.A. Bertulani, P. Danielewicz, Introduction to Nuclear Reactions, IOP
Publishing (CRC), Bristol, UK, 2004 (Chapter 8).

\bibitem{FV77} V. Franco and G.K. Varma, Phys. Rev. C 15, 1375 (1977).

\bibitem{JM73} G. Jacob and Th. A. J. Maris, Rev. Mod. Phys. 38, 121 (1973)

\bibitem{SC86} C.Samanta, N.S.Chant, P.G.Roos, A.Nadasen, J.Wesick  and A.A.Cowley,  Phys. Rev. C 34, 1610 (1986).

\bibitem{Krei95} G. Krein, Th. A. J. Maris, B. B. Rodrigues and E. Veit, Phys. Rev. C 51, 2646 (1995).

\end{thebibliography}
\end{document}